\documentclass[a4paper,11pt]{article}
\pdfoutput=1 % if your are submitting a pdflatex (i.e. if you have
\usepackage{jinstpub} % for details on the use of the package, please
\usepackage{epsfig}
\usepackage{color}

\title{\boldmath Neutron irradiation test of Hamamatsu, SensL and AdvanSiD UV-extended SiPMs}

\author[a]{M.~Cordelli}
\author[a,e,2]{E.~Diociaiuti,\note{Corresponding author.}}
\author[a,b,1]{R.~Donghia,\note{Corresponding author.}}
\author[c]{A.~Ferrari}
\author[a]{S.~Miscetti}
\author[c]{S.~M\"uller} 
\author[a,f]{I.~Sarra}

\affiliation[a]{Laboratori Nazionali di Frascati dell'INFN, Frascati, Italy}
\affiliation[e]{Universit\`a degli Studi Tor Vergata, Rome, Italy}
\affiliation[b]{Universit\`a degli Studi Roma Tre, Rome, Italy}
\affiliation[c]{HZDR - Helmholtz-Zentrum Dresden-Rossendorf, Dresden, Germany}
\affiliation[f]{Universit\`a degli Studi Guglielmo Marconi, Rome, Italy}

\emailAdd{rdonghia@lnf.infn.it, eleonora.diociaiuti@lnf.infn.it}

\abstract{
In this paper, we report the measurement of the neutron radiation hardness of custom Silicon Photomultipliers arrays (SiPMs) manufactured by 
three companies: Hamamatsu (Japan), AdvanSiD (Italy) and SensL (Ireland). 
These custom SiPMs consist of a $2\times~3$ array of $6~\times~6$~mm$^2$ 
monolithic cells with pixel sizes of respectively $50~\mu$m (Hamamatsu and SensL) and $30~\mu$m (AdvanSid).

 A sample from each vendor has been exposed to neutrons generated by the Elbe Positron Source 
 facility (Dresden), up to a total fluence of $\sim 8.5 \times 10^{11}$ n$_{1 \rm{MeV}}$/cm$^2$.
Test results show that the dark current increases almost linearly with the neutron fluence.

The room temperature annealing  was quantified by measuring the dark current two months after the irradiation test.
The dependence of the dark current on the device temperature and on the applied bias have been also evaluated.
}


\arxivnumber{1710.06239} % only if you have one

\begin{document}
\maketitle
\flushbottom

\section{Introduction}
\label{sec:intro}
Silicon Photomultipliers (SiPMs) are semiconductor photon-counting 
devices composed by a matrix of Avalanches Photo Diodes (APDs) operating in Geiger mode.
Their working point is few volts above the breakdown voltage and their output signal is the 
sum of the charge produced by each APD pixel fired by a photon.
The active area and the shape of a SiPM, as well as the dimension of the pixels, can often be customised 
according to the user's needs~\cite{sipm}.

To be used  in High Energy Physics, SiPMs have to withstand 
many years of operation in harsh radiation environments where both
ionising and non-ionising dose are delivered. On reference ~\cite{calor}, we have
tested Hamamatsu~\cite{ham} SiPMs up to 200 Gy using a $^{60}$Co source, observing
negligible effects on gain and leakage current. 
In this paper, we have investigated the damage induced by neutrons 
on SiPMs manufactured by three companies: 
Hamamatsu (Japan), AdvanSid (Italy)~\cite{adv} and SensL (Ireland)~\cite{sen}. 
The SiPMs under test were custom-arrays, each one  constituted by a  $2\times~3$ array of $6~\times~6$~mm$^2$ monolithic cells with 
pixels size of 50 $\mu$m for Hamamatsu and SensL, and 30 $\mu$m for AdvanSid.
%pixel sizes of respectively $50~\mu$m (Hamamatsu and SensL) and $30~\mu$m (AdvanSid).
The tested SiPMs were designed with a thermal package having a thermal resistance below 
$7 \times 10^{-4}$, in order to improve their dissipation capabilities with respect to the standard models
and allow better cooling.
%The three firms provided us SiPMs with thermal packages having resistance lower than 
%$7 \times 10^{-4}$ kW/m$^2$, in order to allow the dissipation of the heat due to the leakage current. 

Three samples, one per vendor, have been exposed to the same neutron fluence 
at the Elbe Positron Source (EPOS) facility of the Helmholtz-Zentrum Dresden-Rossendorf 
(HZDR), Dresden  in March 2017~\cite{hzdr}.

The main effects of  radiation on Silicon are: (i) the bulk damage, caused by the displacement of crystal atoms,
 and (ii) the surface damage, which includes all effects in the covering dielectrics and in the interface region \cite{articolo_radiation}\cite{art_rad2}.
Both types of damage cause an increase of leakage current and noise, change in material resistivity, 
reduction in the amount of collected charge due to the charge carrier trapping 
mechanism and decrease of the carrier's mobility and their lifetime.
In out test,  both type of damage have not been distinguished. 

\section{The EPOS - HZDR Facility}
%%%%%%%%%%%%%%%%%%
\begin{figure}[h!]
\centering
     \includegraphics[width=0.5\textwidth]{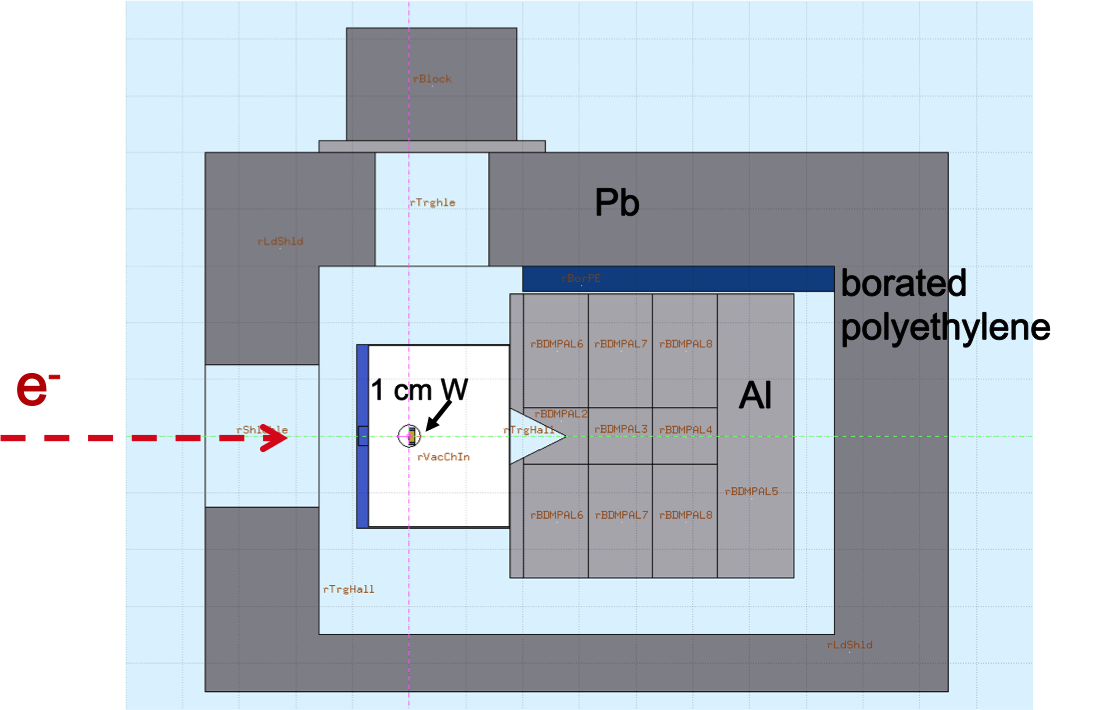} 
% \caption{Sketch of the EPOS facility and of the shielding around the Tungsten target.}
 \caption{Sketch of the EPOS facility and of the lead (borated polyethylene) shielding in grey (dark blue) around the Tungsten target.}
  \label{fig:EPO}
\end{figure}
In the EPOS facility,  a 30 MeV electron beam, of  O(100 $\mu$A)
current, interacts with 1~cm thick Tungsten target and becomes a source of photon
and neutrons~\cite{hzdr_epo}. In Figure \ref{fig:EPO}, the picture of the target with the surrounding 
shielding of Lead and Borated Polyethylene is shown.

\begin{figure}[!b]
  \begin{center}
    \begin{tabular}{cc}
    \includegraphics[width=.5\columnwidth]{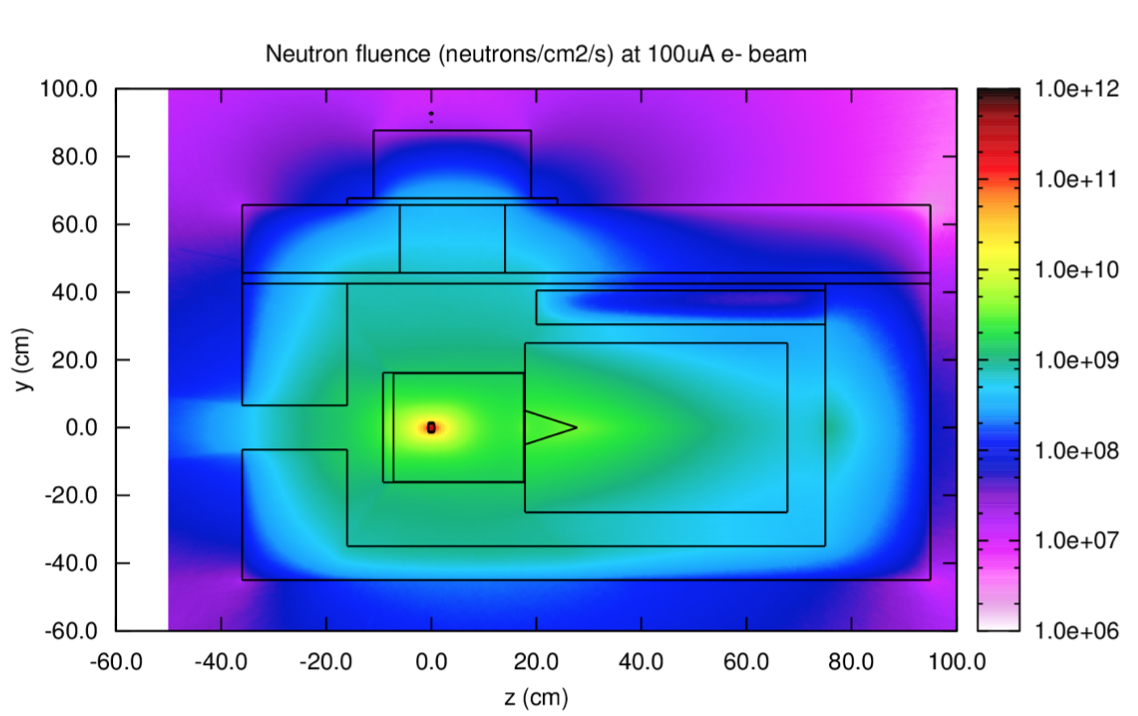} &
\includegraphics[width=.5\columnwidth]{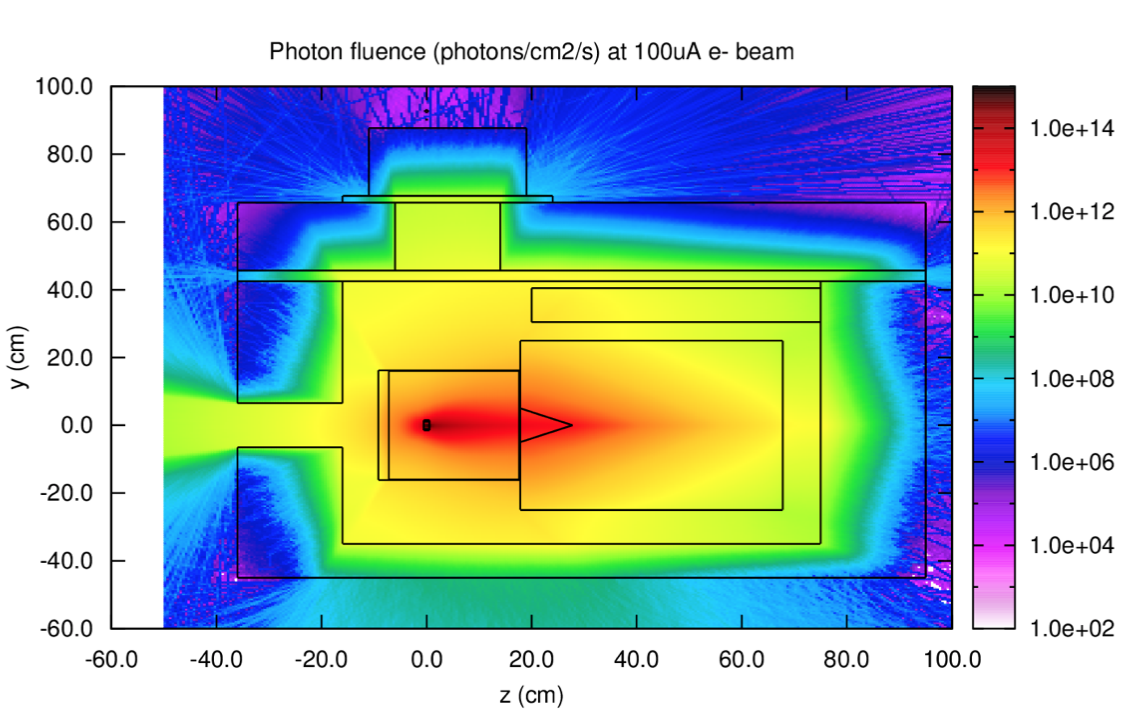}
     \end{tabular}
  \end{center}
  \caption{Distribution of the neutron (left) and of photon (right) fluences as a function of the position around the shielding for an electron current of 100 $\mu$A.}
  \label{fig:EPOS_n}
\end{figure}

In Figure~\ref{fig:EPOS_n} the distribution of the equivalent expected dose for neutrons
is shown on left, while on right the distribution of the expected dose due to photons is presented. 
%Comparing the dose between the two distributions, we have optimised the position
%where to locate the devices under test, i.e. on the top of the shielding roof. 
The devices under test were located on top of the shielding roof, where the expected 
photons contribution in dose is negligible. 
The  ratio between the expected photon and neutron induced doses in this region 
is of $\sim 10^{-4}$.

\begin{figure}[!h]
\begin{center}
\begin{tabular}{cc}
\includegraphics[width=0.45\textwidth]{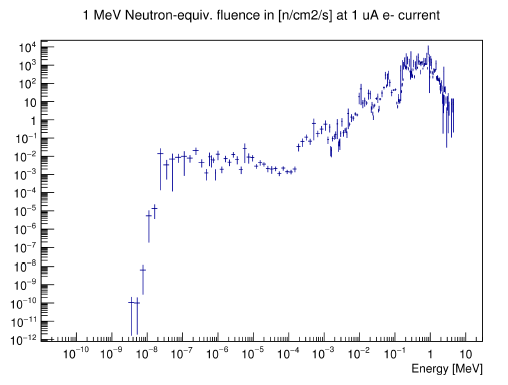} &
\includegraphics[width=0.464\textwidth]{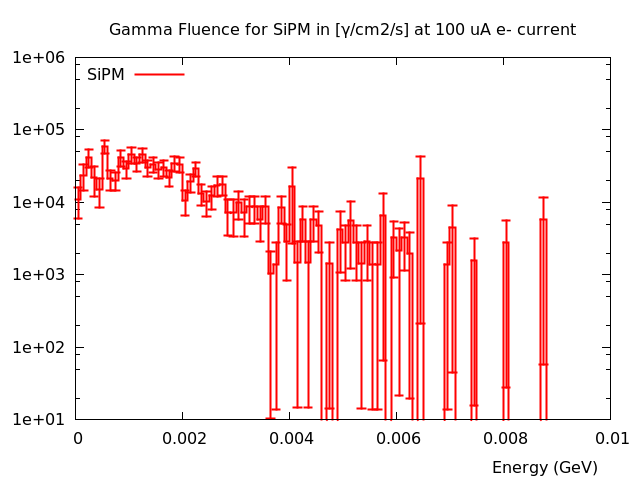} 
\end{tabular}
  \caption{1 MeV equivalent neutron (left) and gamma (right) spectra for an electron current of 1~$\mu$A at the SiPMs position.}
  \label{fig:n_spectr}
  \end{center}
\end{figure}

During the irradiation period, that lasted $\sim$~29~hours, we have continuously recorded the beam
current in order to monitor the flux intensity along the data taking time. The neutron fluence has been
estimated by a full FLUKA simulation~\cite{fluka} and has been scaled to a 1 MeV equivalent neutron 
damage on Silicon as a function of the kinetic energy of the simulated neutrons. The 
spectrum (Figure~\ref{fig:n_spectr}, left) is well centred around 1 MeV. 
Most of the photon produced have an energy below 3~MeV, as shown in Figure~\ref{fig:n_spectr} (right), so they are well shielded by lead. 

%%%%%%%%%%%%%%%%%%%%%
\section{Experimental setup and measurements}
\begin{figure}[!b]
  \begin{center}
    \begin{tabular}{cc}
    \includegraphics[width=.5\columnwidth]{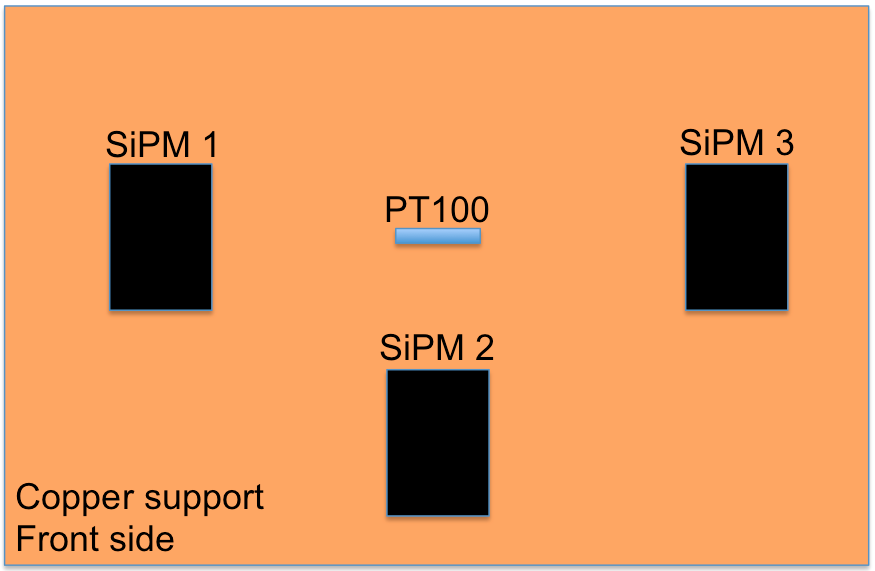} &
\includegraphics[width=.2\columnwidth]{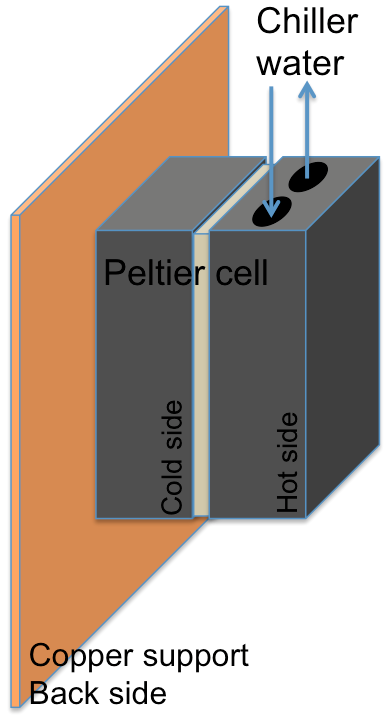}
     \end{tabular}
  \end{center}
  \caption{Front (left) and back (right) sketch of the experimental setup used for SiPMs irradiation test at EPOS facility of HZDR.}
  \label{fig:sketch_dresda}
\end{figure}

%%%%%%%%%%%%%%%%%%%%%
The three samples, one per vendor, have been exposed simultaneously to the same neutron flux.
The sketch of the experimental setup is shown in Figure~\ref{fig:sketch_dresda}, where the
details of the SiPM support are shown.

\begin{figure}[!t]
\begin{center}
    \includegraphics[width=.4\columnwidth]{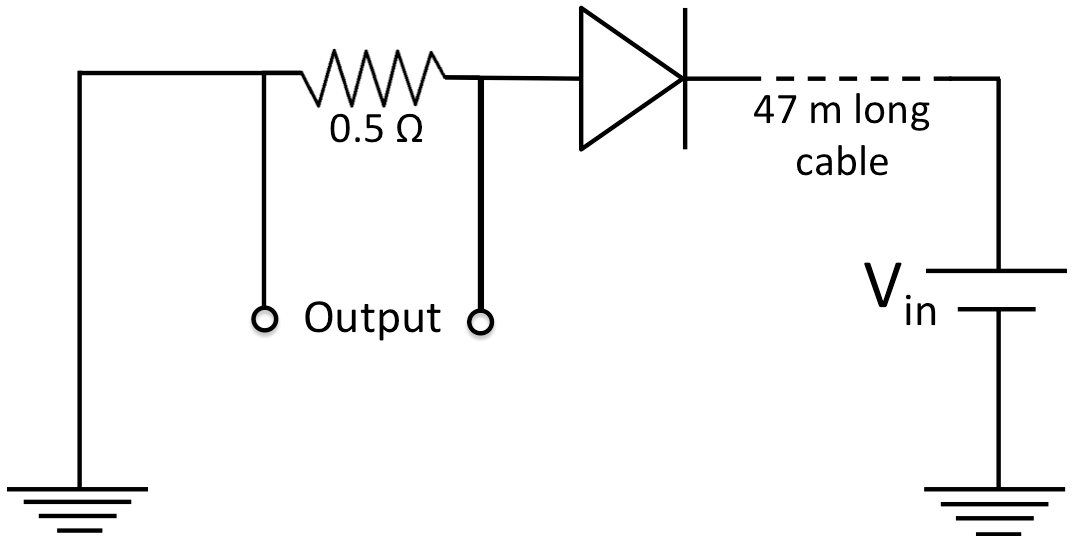} 
\caption{Scheme of the circuit used to bias and acquire the voltage output of the SiPM cell under test.}
\label{circuito}
\end{center}
\end{figure}

The sketch of the used readout circuit  is shown in Figure~\ref{circuito}. 
To evaluate the dark current, the acquired output voltage has been divided by the value (0.5~$\Omega$) of the load resistor. 

To maintain the SiPMs' temperature as stable as possible, the sensors under test have been plugged on a copper support, which 
has been mounted on the cold side of a Peltier cell. In order to assure a good thermal contact, a thermal paste has been used to 
couple  the sensor to the copper plate. The Peltier cell hot side was connected to a very stable chiller system, with water coolant,
running at (18.3 $\pm$~0.1~) $^\circ$C. This allowed us to maintain the other side at around 20$^\circ$~C. 
To monitor the SiPMs temperature along the run, a PT100 resistance thermometer have been plugged in the middle of the SiPMs in the copper support. 
This resistance temperature detector is made by a ceramic material with an accurate resistance-temperature relationship.%, which is used to provide an indication of temperature.

\begin{figure}[!h]
  \begin{center}
    \begin{tabular}{cc}
    \includegraphics[width=.38\columnwidth]{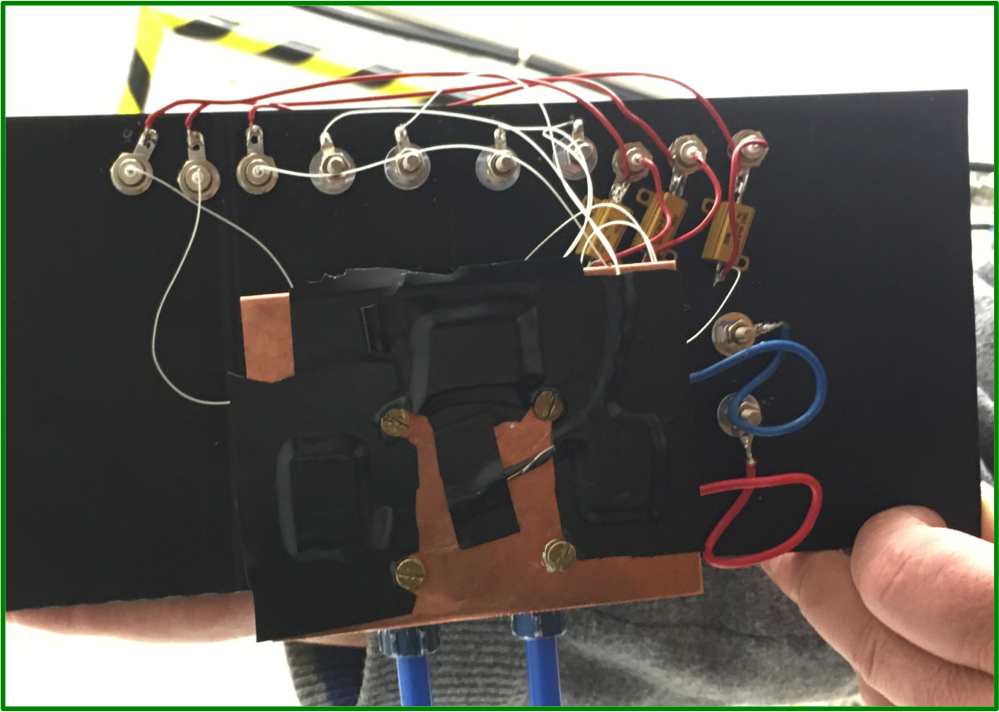} &
\includegraphics[width=.484\columnwidth]{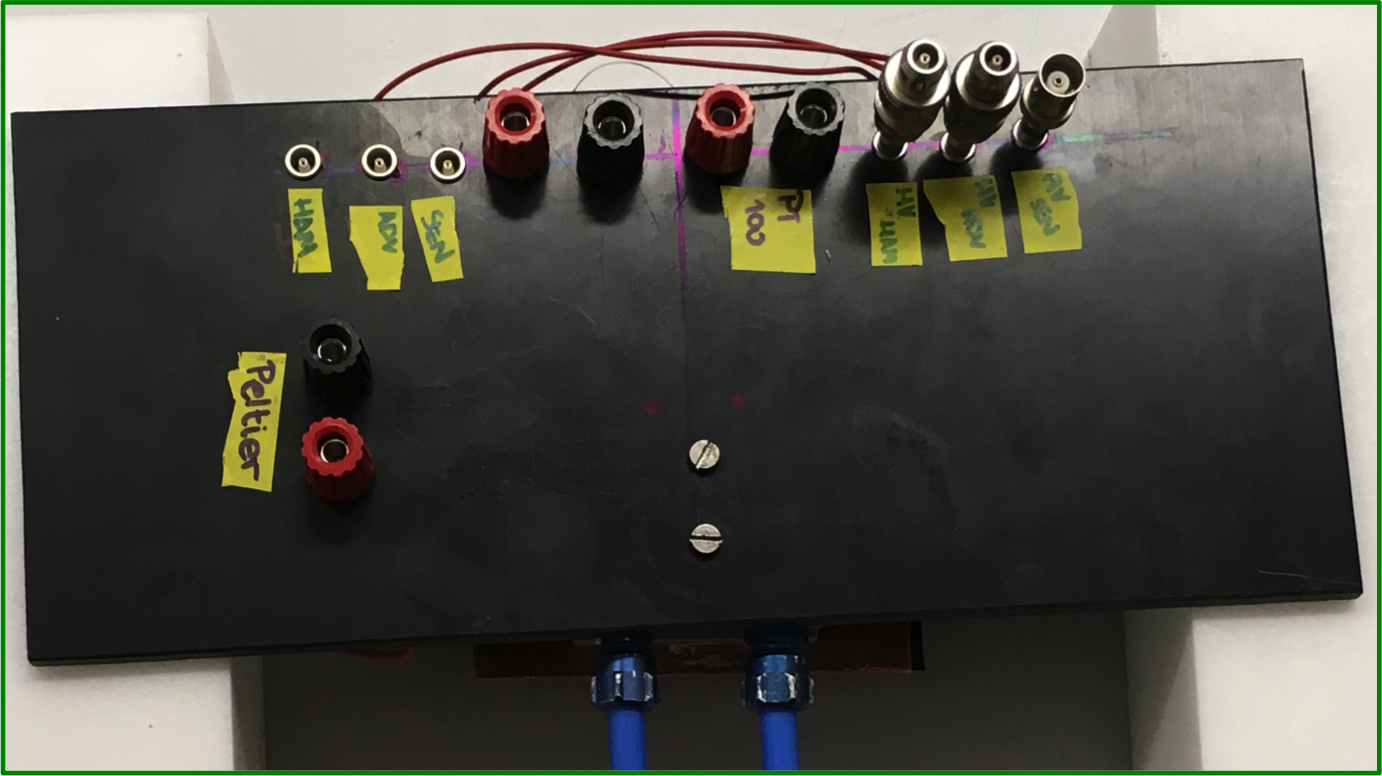}
     \end{tabular}
  \end{center}
  \caption{Pictures of the front (left) and back (right) side  of the experimental setup used for 
  SiPMs irradiation test at EPOS facility of HZDR. The three  SiPMs under test are covered by 
  black tape to make them light tight and to not disturb the measurement of dark current.}
  \label{fig:setup_dresda}
\end{figure}

The SiPM output voltages and the PT100 temperature were acquired by an Agilent 34972A LXI Data Acquisition / Data Logger Switch Unit every 10~s~\cite{datalogger}. Pictures of the experimental setup are shown in Figure~\ref{fig:setup_dresda}.

\begin{figure}[!h]
\centering
\includegraphics[width=0.6\textwidth]{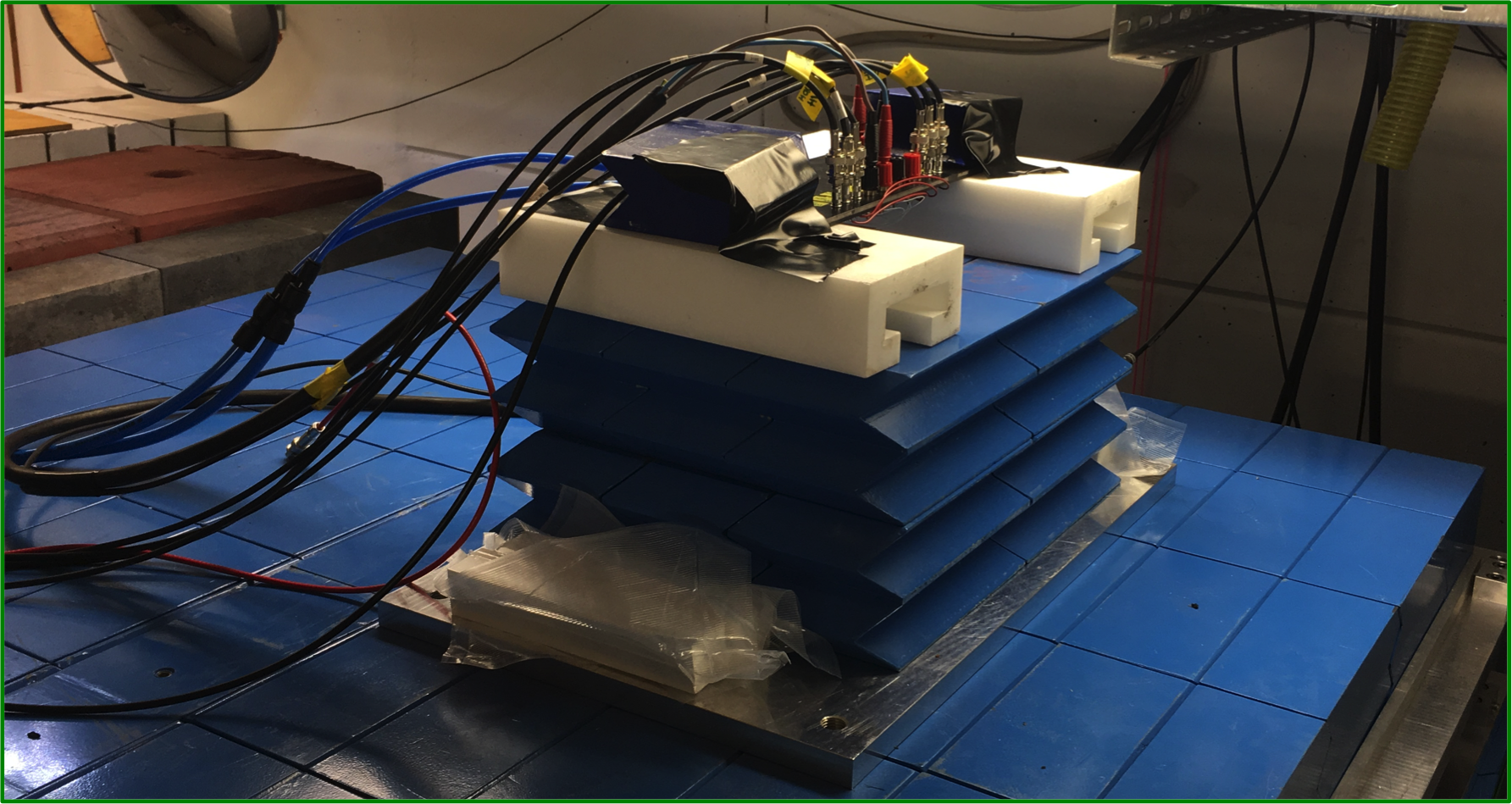}
\caption{Experimental setup mounted in EPOS on top of the shielding. The SiPM surface is located
at 2.2~cm from the neutron  shielding (blue lead blocks) and 90~cm far away from the neutron source.}
\label{fig:source_dr}
\end{figure}

The SiPM support has been placed on the top of the shielding at $\sim$ 90~cm 
far away from the neutron source, with the sensors active area positioned perpendicular 
to the incoming neutrons, as shown in Figure~\ref{fig:source_dr}. 
For each SiPM, only one out of six cells has been biased at the operating voltage, while the other
five cells were not biased. During the irradiation, we have continuously measured the  current drawn
by the biased cells.

\begin{figure}[!h]
  \begin{center}
    \begin{tabular}{cc}
    \includegraphics[width=.5\columnwidth]{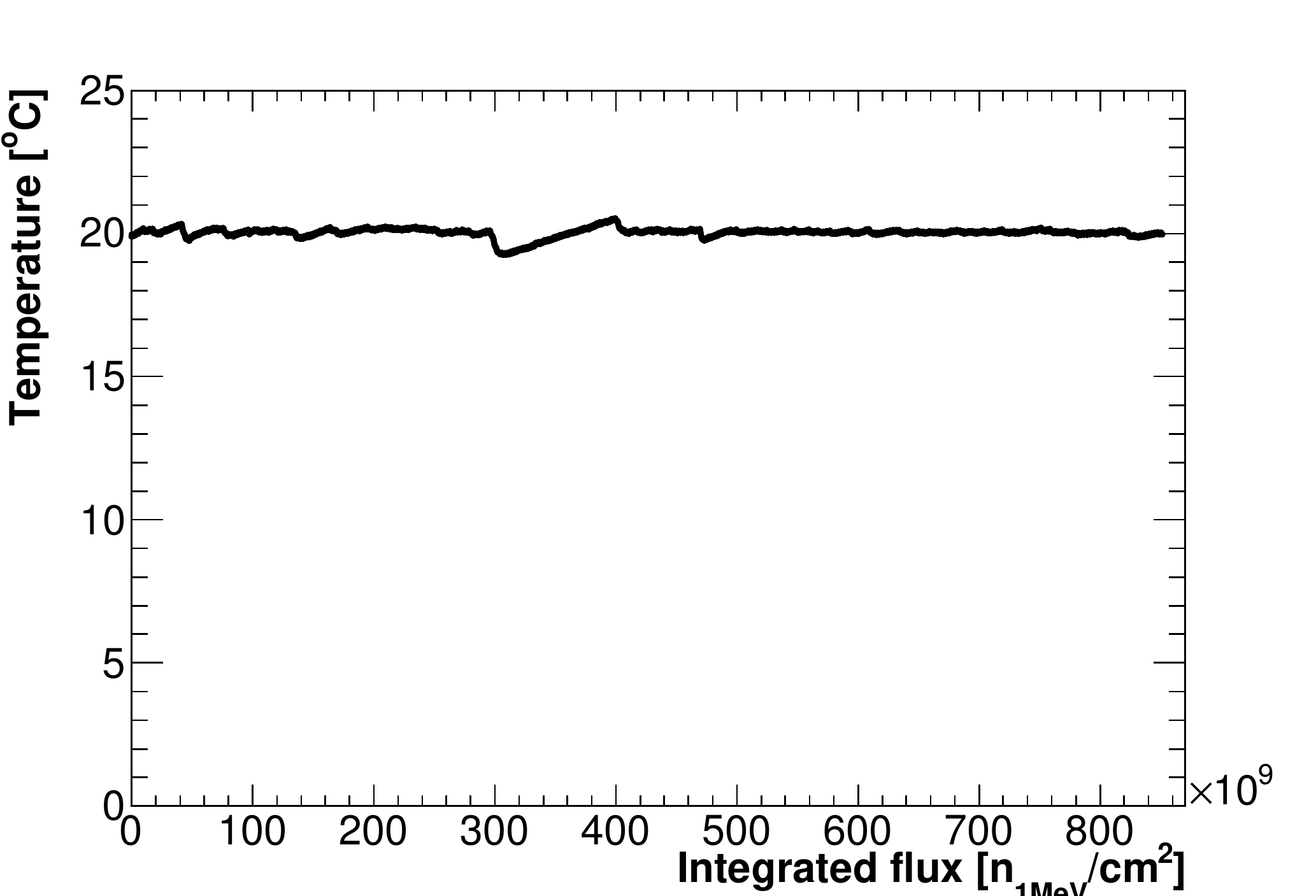} &
\includegraphics[width=.5\columnwidth]{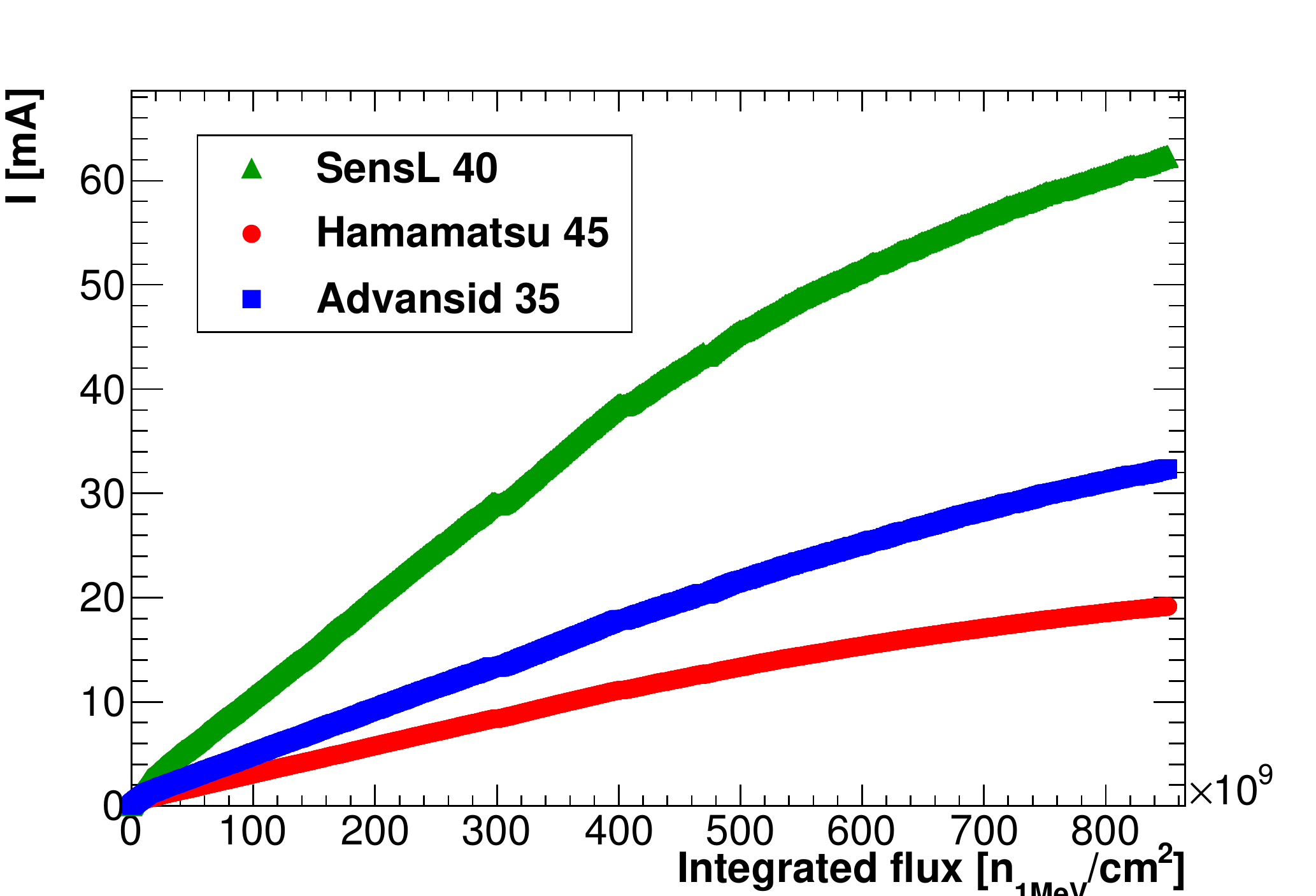}
     \end{tabular}
  \end{center}
  \caption{Left: temperature behaviour of the SiPM copper support as a function of the integrated neutron flux.
  Right: Vendor cells dark current as a function of the integrated neutron flux, delivered in $\sim$~29 hours.}
  \label{fig:t_dresda}
\end{figure}

As shown in Figure~\ref{fig:t_dresda} (left), the Peltier cell and the chiller system were 
performing well and  were able to maintain the copper temperature stable around 20~$^\circ$C.

\begin{figure}[!b]
  \begin{center}
      \epsfig{file=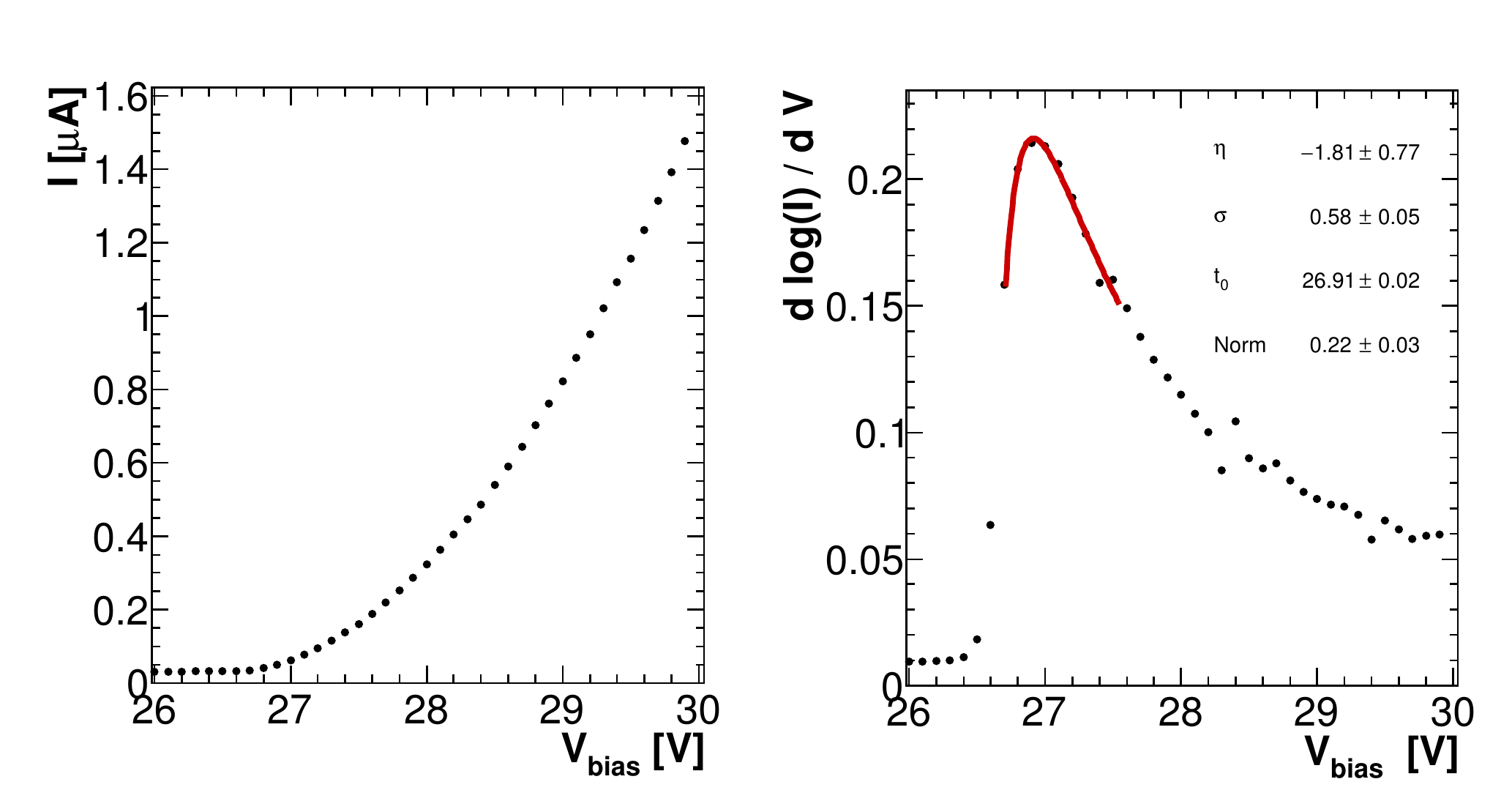,width=.7\textwidth}
  \end{center}
\caption{Dark current logarithmic derivative as function of the SiPM bias voltage. 
The red line represents the log-gaussian fit used to extract the breakdown voltage 
as the bias voltage maximizing the LD value.}
\label{fig:I_vs_V}
\end{figure}

%%%%%%%%%%%%%
%\subsection{Optimization of the operational voltage}
%%%%%%%%%%%%%
The operational voltage (V$_{op}$), of each cell of the SiPMs under test, has been  determined in our 
laboratory in Frascati and has been set at 3~V over the breakdown voltage ($V_{br}$). The $V_{br}$  
has been evaluated with the method of maximization of the current Logarithmic Derivative (LD), that is d(log I)/dV~\cite{vbr}.
To perform this measurement, the SiPMs have been inserted inside a 
darkened black box. The dark current as a function of the bias voltage has been recorded 
using a Keithley pico-ammeter. %, through an automatic acquisition program, that could 
%read the voltage and current information with a programmable voltage step.
An example of the current LD as a function of the bias voltage is shown in Figure~\ref{fig:I_vs_V}.

%%%%%%%%%%%%%
%\section{Results}
%%%%%%%%%%%%%
The total neutron fluence absorbed by the SiPMs at the end of the 29 hours long test was estimated to
be  $\sim~8.5~\times~10^{11}$~n/cm$^2$.
%A full Fluka simulation estimated that the total 1 MeV neutron fluence absorbed 
%by the SiPMs in the 29 hours of run corresponds to $\sim~8.5~\times~10^{11}$~n/cm$^2$.

In Figure~\ref{fig:t_dresda} (right), the measured current of the biased cells 
as a function of the integrated flux are reported for the three tested SiPMs. 
%A clear linear increase of the current as a function of the integrated flux is observed.
To summarise the measurements performed at EPOS, the temperature, 
V$_{op}$ and the dark current (I$_d$) at the end of the irradiation 
are reported in Table~\ref{tab:dresda} for each SiPM.

\begin{table}[h!]
\centering
\begin{tabular}{|c|c|c|c|c|}
\hline
SiPM                 &            T        & V$_{op}$ & I$_d$ & I$_d$  \\ 
vendor               &                      &                &           & (after $\sim2$~months) \\
                          &    [$^\circ$C] &   [V]        & [mA]    & [mA]  \\ \hline
AdvanSiD 35     &          20        & 29.9       &   32.4   &   19.1\\ 
Hamamatsu 45 &          20        & 54.7        & 19.5     &  10.0\\ 
SenSL 40           &          20        & 27.9        & 62        &  38.8\\ \hline
\end{tabular}
\caption{Temperature (T), bias voltage (V$_{op}$) and  dark current ( I$_d$) for each vendor SiPM cell tested at EPOS, at the end of
the irradiation period. Total fluence delivered was of  $\sim 8.5 \times 10^{11}$n$_{\rm{1~MeV}}/$cm$^2$.}
\label{tab:dresda}
\end{table}

The current dependance on neutron fluence is well represented by a linear behaviour up to about 15~mA. 
%The behaviour of the current as a function of the neutron fluence is well represented by a linear 
%$\sim~3\times10^{11}$~n/cm$^{2}$, thus suggesting a negligible gain drop. 
%\textcolor{red}{The deviation from  a linear dependence at higher fluences is  due to the voltage drop on the 47~m long RG58 cables, that corresponds to an additional static load of $\sim$~3.1~$\Omega$. }
%Indeed we measured a decrease in current of a factor 3 lowering the bias voltage of 1~V. 
%This means at 15 mA, the bias drop is about 46.5~V, which increase with the current. 

At the end of the irradiation period, we have also measured the currents of the unbiased cells 
that resulted to be few \% larger than the biased ones. 
The dark current increase of the Hamamatsu SiPM looks much smaller
than the Advansid and Sensl ones.

%\newpage

\section{Current measurement after $\sim 2$ months of annealing}

\begin{figure}[!h]
  \begin{center}
    \includegraphics[width=.7\columnwidth]{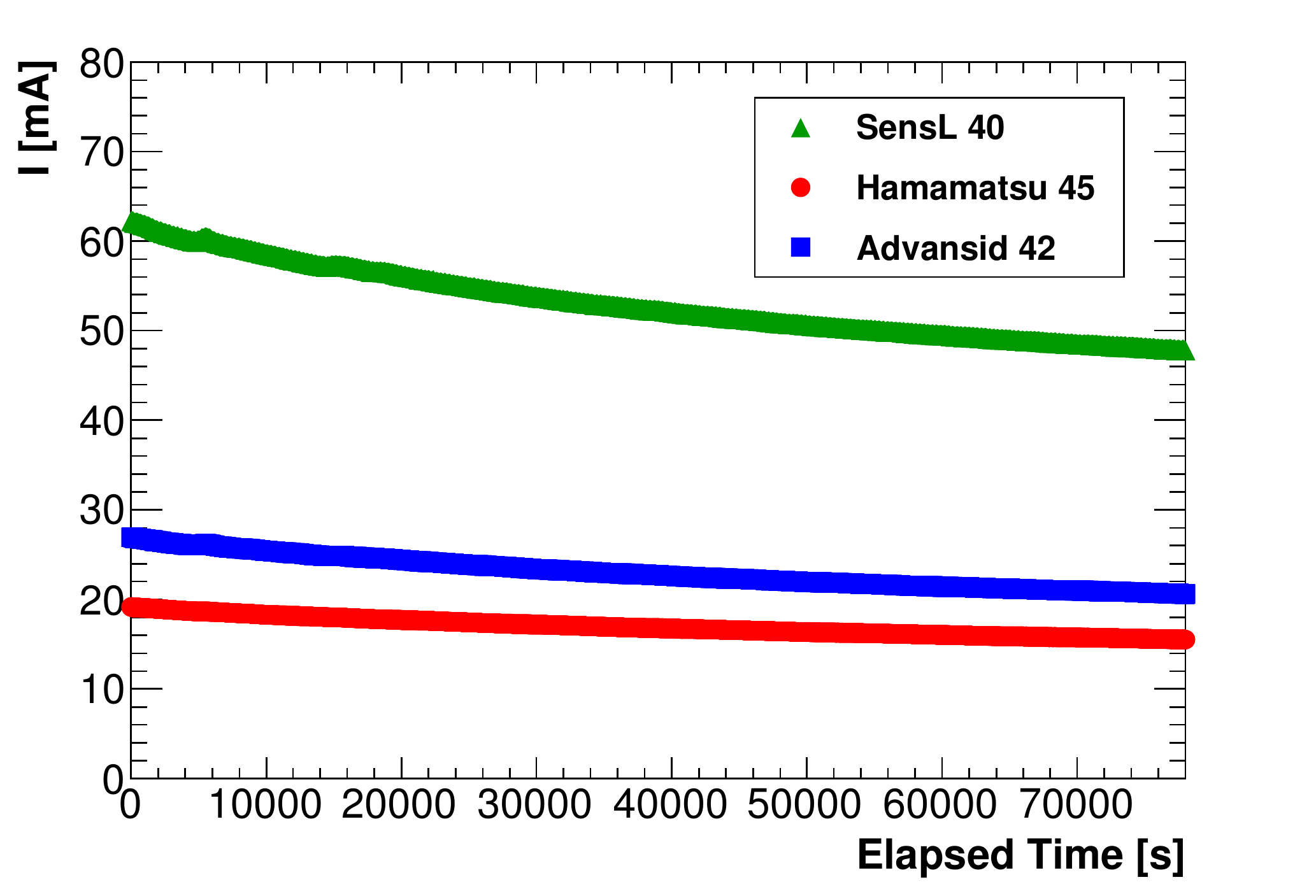} 
  \end{center}
  \caption{Vendors cells dark current as a function of the elapsed time after the end of the neutron fluence exposure.}
  \label{fig:anneal}
\end{figure}

We have continued to measure the current for about 22 hours after the end of the beam exposure, 
as shown in Figure~\ref{fig:anneal}. 
 
A current decrease due to room temperature annealing is clearly visible.
Indeed, two months after the end of the irradiation test, a leakage current reduction of about 50\% has been observed
as shown in Table~\ref{tab:dresda} (last column).
The SiPM cells biased during  irradiation have been tested to determine the dark current variation with  respect 
to a change in temperature and bias voltage.  
The experimental setup previously described has been inserted in a light-tight, insulated box, fluxed with 
nitrogen, in order to decrease the temperature without reaching the dew point.

The dark current (I$_{\rm d}$)  has been measured by first  varying the temperature of the SiPMs 
and then varying the operational voltage at fixed temperature.

The SiPMs temperature has been decreased from 20~$^\circ$C down to -5$~^\circ$C. 
In order to keep unchanged the operational point while varying the temperature, 
the bias voltage has been decreased by $0.1 \%$ per degree, as from technical specifications. 
This voltage adjustment has been cross-checked by measuring, for each SiPM, the breakdown voltage at 20, 
10 and 0~$^\circ$C.
  
\begin{figure}[!h]
  \begin{center}
    \begin{tabular}{cc}
    \includegraphics[width=.5\columnwidth]{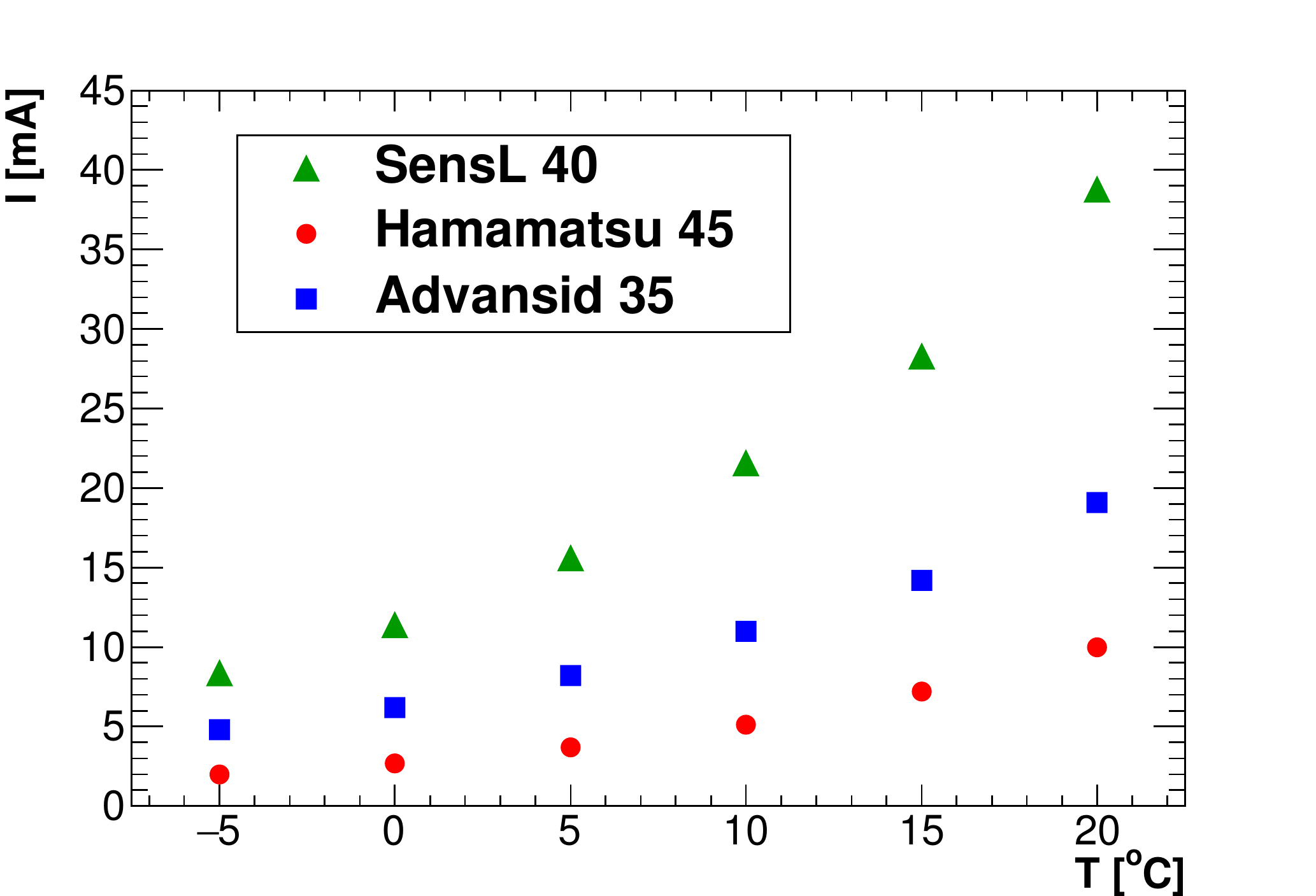} &
\includegraphics[width=.5\columnwidth]{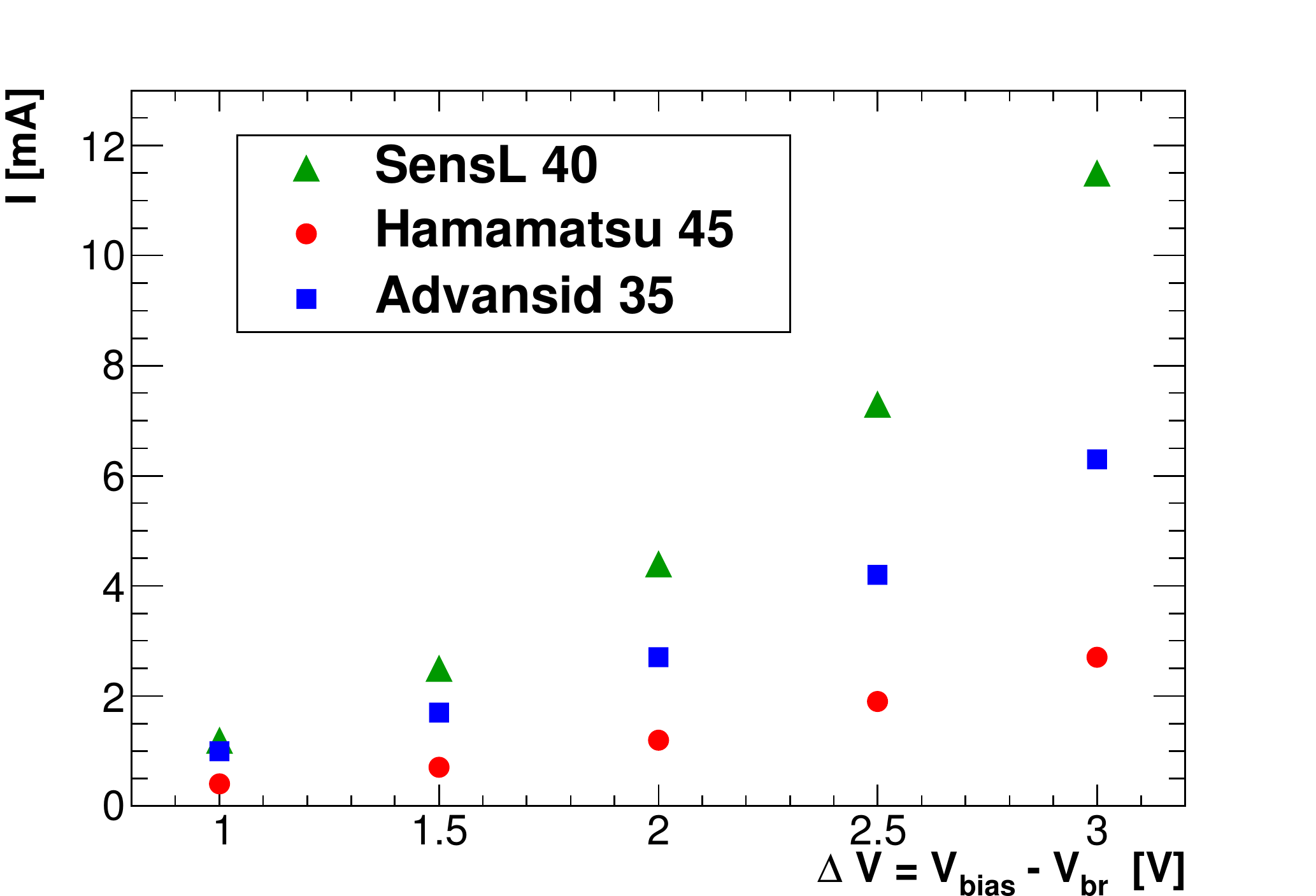}
     \end{tabular}
  \end{center}
  \caption{Left: Vendors cell current as a function of the temperature at the operational voltage. 
    Right: Vendor cells current as a function of the $\Delta V $. Temperature was kept stable at 0~$^\circ$ C.}
  \label{fig:I_Temp}
\end{figure}

The variation of the measured I$_{\rm d}$  with respect to temperature is reported in \figurename~\ref{fig:I_Temp} (left).
A decrease of 10~$^\circ$C in the SiPM temperature corresponds to a decrease of about 50\% in I$_{\rm d}$.
After completing this first study, the SiPM temperature has been fixed at 0~$^\circ$C and 
the current has been acquired at different bias voltages. Results are reported in 
\figurename~\ref{fig:I_Temp} (right), as a function of the difference between the bias and the breakdown voltage ($\Delta V $).
Also in this case, a reduction of the working points quickly diminishes  the drawn dark current.

%%%%%%%%%%%%%
\section{Conclusions}
%%%%%%%%%%%%%
The neutron radiation damage for  three custom SiPMs from different vendors (AdvanSiD, Hamamatsu and SensL) 
have been determined using neutrons generated at the EPOS facility of HZDR, Dresden.
The SiPMs used consisted of the parallel configuration of two series of three $6\times6$ mm$^2$ cells.
The dark current of the SiPMs increases linearly as a function of the 
1~MeV equivalent neutron fluence up to 15~mA, while a deviation from linearity is observed at larger dark currents, 
as due to the additional voltage drop in the 47 m long cable.
% $\sim~3\times10^{11}$~n/cm$^{2}$ while a deviation from linearity is observed at larger neutron fluences, as due to the additional voltage drop in the 47 m long cable.

The dark current increase in the SensL and AdvanSiD sensors resulted to be much faster than that
of Hamamatsu SiPMs. At the end of the irradiation period, the dark current values
were  of 62 mA, 32.4 mA and 19.5 mA for SensL, AdvanSiD and Hamamatsu respectively.
After two months of annealing at room temperature the dark current values  decreased by a factor  of two.

We have also been observing that lowering the sensor temperature of 10~$^\circ$C, 
as well as decreasing the bias voltage of 1~V/cell,  corresponds to a reduction of an additional
factor of two in the dark current.

%
%
%A temperature dependent self-annealing effect is observed.
%The dark current of AdvanSiD and (much more) SensL  SiPMs rise faster than Hamamatsu devices when they are exposed to an intense neutron flux.

\acknowledgments
%\section*{Acknowledgments}
This work was supported by  the EU Horizon 2020 Research and Innovation Program 
under the Marie Sklodowska-Curie Grant Agreement No.690385. 
The authors are grateful to many people for the successful realisation of the
tests. In particular, we thank all HZDR-EPOS staff, for providing a necessary and well equipped facility,
and all the LNF technicians helping for making this test a reality.

% We suggest to always provide author, title and journal data:
% in short all the informations that clearly identify a document.

%\begin{thebibliography}{99}

%\bibitem{a}
%Author, \emph{Title}, \emph{J. Abbrev.} {\bf vol} (year) pg.

% Please avoid comments such as "For a review'', "For some examples",
% "and references therein" or move them in the text. In general,
% please leave only references in the bibliography and move all
% accessory text in footnotes.

% Also, please have only one work for each \bibitem.
%\end{thebibliography}

\end{document}